\documentclass[12pt,preprint]{aastex}

\newcommand{\eb}{\begin{equation}}
\newcommand{\ee}{\end{equation}}

\newcounter{arabnum}
\newcommand{\befigcap}{\begin{list}{ {\bf Figure \arabic{arabnum} } %
{ \usecounter{arabnum}} } }
\newcommand{\enfigcap}{\end{list}}

\newcommand{\bequo}{\begin{quotation}}
\newcommand{\enquo}{\end{quotation}}
\newcommand{\bverse}{\begin{verse}}
\newcommand{\everse}{\end{verse}}
\newcommand{\beit}{\begin{itemize}}
\newcommand{\enit}{\end{itemize}}
\newcommand{\been}{\begin{enumerate}}
\newcommand{\enen}{\end{enumerate}}
\newcommand{\ecen}{\end{center}}
\newcommand{\bcen}{\begin{center}}
\newcommand{\begeq}{\begin{equation}}
\newcommand{\eneq}{\end{equation}}
\newcommand{\befig}{\begin{figure}}
\newcommand{\enfig}{\end{figure}}

\begin{document}


\shorttitle{Census of SDSS Objects from {\it ClassX}}
\shortauthors{Suchkov, Hanisch \& Margon}

\title{A Census of Object Types and Redshift Estimates in the SDSS 
Photometric Catalog from a Trained Decision-Tree Classifier}
\author{A. A. Suchkov, R. J. Hanisch, and Bruce Margon} 
\affil{Space Telescope Science Institute\footnote{Operated by AURA Inc., 
under contract with NASA},
3700 San Martin Dr., Baltimore, MD 21218
}

\begin{abstract}

\medskip

We have applied ClassX, an oblique decision tree classifier optimized for
astronomical analysis, to the homogeneous multicolor imaging data base of
the Sloan Digital Sky Survey (SDSS), training the software on subsets of
SDSS objects whose nature is precisely known via spectroscopy.  We find
that the software, using photometric data only, correctly classifies a
very large fraction of the objects with existing SDSS spectra, both
stellar and extragalactic. ClassX also accurately predicts the redshifts
of both normal and active galaxies in SDSS.  To illustrate ClassX
applications in SDSS research, we (a) derive the object content of the
SDSS DR2 photometric catalog and (b)  provide a sample catalog of resolved
SDSS objects that contains a large number of candidate AGN galaxies,
27,000, along with 63,000 candidate normal galaxies at magnitudes
substantially fainter than typical magnitudes of SDSS spectroscopic
objects. The surface density of AGN selected by ClassX to $i\sim 19$ is in
agreement with that quoted by SDSS. When ClassX is applied to the
photometric data fainter than the SDSS spectroscopic limit, the inferred
surface density of AGN rises sharply, as expected.  The ability of the
classifier to accurately constrain the redshifts of huge numbers
(ultimately $\sim\! 10^7$) of active galaxies in the photometric data base
promises new insights into fundamental issues of AGN research, such as the
evolution of the AGN luminosity function with cosmic time, the
starburst--AGN connection, and AGN--galactic morphology relationships.

\end{abstract}


\section{Introduction \label{s-intro}}

The Sloan Digital Sky Survey (SDSS; York et al.\ 2000) Data Release 2 
(DR2)\footnote{http://www.sdss.org/dr2} 
photometric (imaging) catalog contains 88 million unique
objects. The DR2 spectroscopic catalog contains 260,000 galaxies, 
35,000 AGN, 35,000 stars of type K and earlier, and 13,000 M and later  
type stars (Abazajian et al.\ 2004a). These numbers are nearly doubled
in the Data Release 3 (DR3; Abazajian et al.\ 2004b). This large amount of  
high quality, homogeneous data creates unique opportunities in many 
fields of current research. The use of new
technologies capable of analyzing very large databases
promises results unachievable with old techniques. 

The SDSS imaging database will eventually contain about 2 billion
objects.  Of these only about 1 million objects will be observed
spectroscopically to obtain source classifications and redshifts.
Obtaining reliable object type and redshift estimates based on SDSS 
photometry is thus an extremely valuable adjunct to the spectroscopic sample.

It has long been known that  multicolor photometry can be used 
for object classification and redshift estimation. 
Colors were used for selection of active galaxy candidates for 
decades (see, e.g., Hartwick \& Shade 1990 for the history of the issue).
However, it was only after the mid-1990s that major developments in 
the generation of very large high-quality photometric surveys prompted 
the creation of powerful classification techniques.  

Wolf, Meisenheimer, \& R\"{o}ser (2001) described  the diversity of 
issues one encounters in the development of classification methods 
and their application to specific surveys. 
They designed and implemented a classification algorithm relying on 
a library of color templates, which allows one to identify stars, normal 
galaxies, and active galaxies in multi-color surveys and estimate 
redshifts of the normal and active galaxies. 
The method was applied to tens of thousand objects from the project 
COMBO-17 (Classifying Objects by Medium-Band Observations in 17 Filters), 
yielding important results on the evolution of the galaxy luminosity
function up to $z=1.2$ (Wolf et al.\ 2003a) and evolution of faint 
AGN at $1 < z < 5$ (Wolf et al.\ 2003b). It was used 
to classify and analyze the COMBO-17 objects in the Chandra Deep 
Field South (CDFS) and to construct a catalog of over 60,000 photometric 
objects in that field (Wolf et al.\ 2004). An enhanced version of this method
was applied to identify  object types and estimate redshifts 
of specific X-ray sources in CDFS and construct a catalog of these
sources (Zheng et al.\ 2004). Brand et al.\ (2005) applied a photometric
redshift technique to determine redshifts of a few thousand red galaxies
in the Bo\"{o}tes field and used them to study the nuclear accretion 
history of the red galaxy population.

Photometric classification and redshift estimation is 
of prime importance for the SDSS project.  
The SDSS photometric system was designed to allow one
to derive redshift estimates from five-band 
photometry (Richards et al.\ 2001a; Budavari et al.\ 2001). 
A detailed discussion of the relationships between the SDSS colors
and redshift is given by Richards et al.\ (2001b). SDSS colors feature 
prominently in the algorithm used to select AGN candidates for 
subsequent SDSS spectroscopy (Richards et al.\ 2002). 
Csabai et al.\ (2003) used a range of photometric techniques to 
estimate redshifts of galaxies in the SDSS Early Data 
Release (EDR) catalog, discussing in detail the caveats and issues 
to be kept in mind as one applies these redshift to statistical studies 
of galaxies. They found that the photometric redshift relation
and the resulting redshift histogram are well matched to existing 
redshift survey. Most importantly, they 
Richards et al.\ (2004) found that the SDSS photometric redshifts are
quite suitable for statistical studies of AGN, yielding 
results in agreement with those from the 2QZ spectroscopic survey
(Croom et al.\ 2004) in cases where the comparison is possible. 
In particular, they indicated that the distribution of 
photometric redshifts of the AGN candidates from the SDSS DR1 
photometric catalog is similar to the redshift distribution of the AGN 
in the  2QZ, and the number counts of the SDSS DR1 AGN
candidates are in agreement with that found from the 2QZ/6QZ surveys.

The basis for the classification of the SDSS photometric database
can be provided by the objects whose nature is precisely  known    
from spectroscopy. Richards et al.\ (2004) developed a classification  
technique in  which the software learns from spectroscopic
objects of known identity to recognize the physical type 
of SDSS photometric objects. 
That technique was applied to select AGN candidates 
from the SDSS Data Release 1 (DR1) photometric catalog and resulted
in a catalog of $\sim\!10^5$ unresolved AGN candidates,  
of which 95\% were estimated to be actual AGN. 
That paper also demonstrated the potential of bulk 
classification of the SDSS data and indicated a wide range of research 
applications.

The next SDSS data release, SDSS DR2, is substantially different 
from both the Early Data Release (Stoughton et al.\ 2002) and the SDSS DR1
(Abazajian et al.\ 2003) both in terms of the number of the cataloged
objects and the quality of the data. The content of the spectroscopic
catalog is determined by the way the photometric objects were selected
for the follow-up spectroscopy. The selection criteria were different
for different object types, meaning that the catalog content cannot be
regarded as a uniform representation of the SDSS imaging database
(see Strauss et al.\ 2002 for the galaxy selection criteria, and
Richards et al.\ 2002 for the AGN selection algorithm).

Unbiased results on the object content of the SDSS imaging database 
can only be obtained using a system capable of classification (as opposed
to selection) of the SDSS photometric data into object types of interest. 
With the large
number of currently available SDSS objects whose identity has been established 
through spectroscopy, it is now possible to employ a very powerful 
technique of supervised classifiers, such as ClassX classifiers 
(McGlynn et al.\ 2004; Suchkov et al.\ 2003).
ClassX has proven to be an efficient system for classification of large
sets of objects from  multiwavelength catalogs. McGlynn et al.\ (2004)
presented a catalog for two hundred thousand ROSAT sources
classified with ClassX into six class categories:  star, white dwarf, 
X-ray binary, AGN, galaxy, and cluster of galaxies.  
ClassX is also efficient in identifying rare, interesting
objects.  Suchkov \& Hanisch (2004a) applied these classifiers 
to search for new Galactic X-ray binaries. They detected a significant 
population of low-luminosity, hard X-ray binaries that have 
interesting
implications for the  origin and the nature of various types of X-ray
binaries and their role in the X-ray properties of galaxies.  With a
classifier that utilizes both X-ray and infrared information to
categorize X-ray sources into eight classes, including three spectral types
of stars, Suchkov \& Hanisch (2004b) found a significant population of
extremely obscured sources with all indications of being nascent,
pre-main-sequence stars deeply embedded in the dense, dusty clouds of 
star formation regions. 

ClassX offers new and efficient ways to identify the physical 
nature of SDSS sources. It complements and substantially expands 
the previous work in the field, and has a strong potential  to become 
an important classification tool for the bulk of the SDSS 
photometric database.  
In this paper we use ClassX to analyze the SDSS DR2 photometric 
catalog, classifying SDSS photometric objects into stars, normal galaxies, 
and active galaxies, 
and determining the most likely redshifts of objects classified 
as normal and active galaxies.  We estimate the content of the catalog
(the fraction of objects of different types) and discuss it as 
a function of limiting magnitude. To further illustrate 
ClassX  application to the SDSS research,
we present a sample catalog containing $9\times10^4$ spatially 
extended SDSS sources (i.e., objects with SDSS morphological type 3)
classified into normal galaxies and resolved AGN galaxies.
We expect that the large number of new candidate AGN galaxies 
easily identified by ClassX in the SDSS photometric catalog  
among the resolved objects would result in new insights into many 
issues of current interest, such as the
starburst--AGN connection and the ISM of active galaxies 
(e.g., Scoville 2003; Scoville et al.\ 2003; Imanishi \& Wada 2004), 
normal and star-forming X-ray galaxies (e.g., Anderson et al.\ 2003;
 Zheng et al.\ 2004; Horschenmeier et al.\ 2005), 
X-ray-bright, optically normal galaxies
(XBONGS; e.g., Comastri et al.\ 2002; Yuan \& Narayan 2004), and the
star formation and mass metallicity relation in the low-redshift
universe (e.g., Wolf et al.\ 2003a, 
Brinchmann et al.\ 2004; Tremonti et al.\ 2004). 

In this paper we describe the application of ClassX to the SDSS DR2
and present initial results regarding the classification of normal galaxies
and AGN---separated into redshift bins---from the photometric catalog.
Subsequent analyses will focus on more complete interpretation of the
results vis-a-vis number counts and AGN evolutionary models.

\section{Data}

\subsection{Samples of Spectroscopic and Photometric Objects
\label{sec-samples}}

The sample of SDSS DR2  spectroscopic objects that we used to build and
validate the ClassX classifier includes four major spectroscopic types,
or classes defined in the SDSS: stars (type K and earlier), galaxies 
(resolved SDSS sources), 
AGN (includes resolved and unresolved AGN objects, often referred to 
in the literature as AGN galaxies and quasars, respectively), and red stars
(type M and later).  Each class in the  sample except for the class red
star contains $2\times 10^4$ objects, a sufficient number for training and
validation without incurring excessive computational cost. Class red star
contains 3,852 objects, which is all that are available for this class in
the SDSS DR2 spectroscopic catalog for the adopted magnitude constraints.
This sample total is thus 63,852 objects.

To probe the SDSS DR2 photometric catalog with ClassX, we created three
samples, each containing $1\times10^5$ SDSS photometric objects.  The
size of the samples was selected such as to keep them manageable but
large enough to be representative of the SDSS DR2 photometric database.
The ``bright'' sample is  limited to the brightness range covering the
bulk of the SDSS DR2 spectroscopic catalog.  In each band, the lower
and upper magnitude limits for this sample are approximately 1~$\sigma$
brighter and 1~$\sigma$ fainter, respectively, than the mean of the
magnitude distribution of the spectroscopic sample, where $\sigma$ is
the standard deviation of the magnitude distribution. 
The second sample, called ``intermediate'',  has the upper magnitude limit
in all bands 1~mag fainter than in the bright sample. 
Finally, the ``faint'' sample is 2~mag fainter than the
bright sample in all bands except for the $u$-band, in which the limit
is the same as in the intermediate sample.
The summary of the sample definitions is given
in Table~\ref{t_photo_samples}.

All three photometric samples are constrained to objects with ``clean'' 
photometry, which excludes objects that are blended and/or  saturated,
objects that potentially are electronic ``ghosts'', 
objects that are affected by cosmic rays,
and ``child'' objects.  The actual database query
for the faint sample is as follows: \\
\\
{\tt
SELECT top 100000  p.dered\_u, p.dered\_g, p.dered\_r, p.dered\_i, p.dered\_z, \\
 p.ra,  p.dec, p.type \\
FROM PhotoObj p \\
WHERE \\
(p.flags \& 0x0000000000000008) = 0 AND  \\
(p.flags \& 0x0000000000040000) = 0 AND  \\
(p.flags \& 0x0000000000000010) = 0 AND  \\
(p.flags \& 0x0000010000000000) = 0 AND  \\
(p.flags \& 0x0100000000000000) = 0 AND  \\
(p.flags \& 0x0200000000000000) = 0 AND  \\
p.u > 17.0 AND p.u < 21.5 AND \\
p.g > 16.0 AND p.g < 21.5 AND \\
p.r > 15.5 AND p.r < 21.0 AND \\
p.i > 15.0 AND p.i < 21.0 AND \\
p.z > 14.5 AND p.z < 21.0 \\
}

Of the total $8.8\times 10^7$ objects in the 
SDSS DR2 photometric catalog, the number of objects satisfying
the constraints for the three samples is $3.8\times 10^6$,  
$6.4\times 10^6$, and  $7.0\times 10^6$, respectively. 
For the faint sample the limiting $z$~magnitude, $z_{lim} = 21.0$, 
is 0.5 mag fainter than the nominal completeness limit of 20.5 given for 
the $z$~band in SDSS DR2. However, the fraction of objects within 
$z = 20.5 - 21.0$ is very small, 0.4\% (because of the constraints
in other bands, especially in the $u$~band). Therefore, 
the respective incompleteness effects should be quite small.

For both the spectroscopic and photometric samples we retrieved 
the dereddened magnitudes (henceforth denoted as $u,g,r,i,$ and $z$;
model magnitudes are not used further in the text, so
this notation should lead to no confusion). Also we retrieved the
morphological (photometric) type, which is 6 for point (unresolved)
 sources and 3 for sources resolved in SDSS imaging.
The spectroscopic sample also includes redshift from spectra, $z_{\rm{sp}}$. 
Morphological information was retrieved from Table {\tt PhotoObjAll},
while the spectroscopic objects were retrieved from Table
{\tt SpecObj}.   

\subsection{Training and Validation Samples \label{training-and-validation}}

The spectroscopic sample described above was split into two equal parts,
in which classes star, AGN, and galaxy are represented by 10,000 objects
per class. Class red star  is represented by 1,000 and 2,852  objects 
in the first and second parts, respectively. The first part is used to 
train the ClassX classifier; we call it the {\it training} sample.
The second part, which we call the  {\it validation} sample, is used for
two purposes. First, it serves as a data source to  validate the
classifier. Second, it is a resource to obtain the coefficients used  to
calculate purity and completeness of class populations derived by ClassX
from the photometric database.  The objects in the validation sample are
{\it not} known to the ClassX classifier, because they are {\it not}
used in classifier training.

\section{Object Type and Redshift Determination with ClassX}

\subsection{ClassX Technique  }

ClassX has been originally developed for automated classification of
X-ray sources (McGlynn et al.\ 2004; Suchkov \& Hanisch 2004a;
 Suchkov et al.\ 2003). It is deployed on the Web as a publicly available 
online system\footnote{http://heasarc.gsfc.nasa.gov/classx}. Through the
Virtual Observatory (VO) protocols\footnote{http://www.ivoa.net/ and
http://www.us-vo.net/}, it collects the data necessary for classification
from the worldwide network of online data archives and performs
classification for a user-submitted list of targets.

ClassX is based on a machine learning technology, wherein {\it supervised}
classifiers are ``trained'' to recognize objects of unknown identity by
``learning'' object class properties from samples of
objects whose type, or class, is known.

A training sample  for ClassX is characterized by a set of classes,
where each class is characterized by the same set of attributes;
the same sets of classes and attributes are used by the classifier
to perform classification of unidentified sources. The result of the
training procedure is a ClassX {\it classifier}, which is a set of 
oblique decision trees (10 decision trees for the classifier used in
this paper). The algorithmic core  of ClassX is  the OC1 system 
of Murthy, Kasif, \& Salzberg (1994). 

In ClassX, each tree independently performs classification, after which
ClassX conducts weighted voting of individual classifications using
the scheme proposed by White et al.\ (2000). Normalized weighted votes
represent, in essence, probabilities for an object to belong to any of the
classes defined by the classifier. The class with the highest probability
is adopted as the class of the object under classification, while the
votes yield the class probability distribution for that object. 
For each input object ClassX reports both the class assignment and 
the class probability distribution.

Within ClassX one can build a variety of classifiers optimized for
specific research goals or just to be used individually and/or in
combination to optimize classification for various object types. Different
classifiers will use different sets of classes and/or different sets of class
attributes. For SDSS photometric catalogs, the attributes can be
various combinations of SDSS magnitudes, colors, and morphological
types, while classes are the SDSS spectroscopic types and other types
of objects isolated in samples of SDSS objects, such as white dwarfs,
carbon stars, etc. Taking advantage of the richness and high quality
of the SDSS database,  one can also introduce less conventional classes,
such as the redshift classes used in this paper.  The latter case is an
example of a conversion of a natural object attribute, redshift, into a set
of classes. This extends ClassX capabilities from mere object classification
into the domain of determination of object properties.

ClassX includes  powerful tools to compute the completeness and purity
of the classification results.  The classifier preference matrix provides
coefficients to calculate completeness and purity  and also to directly
determine the nature of contaminants within a given class.  Such information
is useful in understanding which parameters are most influential for the
classifier and in optimizing classifiers for particular object types.

\subsection{ClassX Classifier}

To classify objects in the SDSS DR2 photometric catalog and obtain
redshift estimates for objects identified as type AGN and type galaxy,
we use a ClassX classifier that was built using the training set described
in \S2.
The  classes recognized by the classifier are stars, red stars, 10
redshift classes derived by splitting the galaxy objects into 10
redshift bins, and 13 redshift classes derived by splitting the AGN
into 13 redshift bins. 

There is a great degree of flexibility in the definition of redshift
classes. If we are interested in the large scale redshift distribution
of AGN, we can define such a class as a redshift bin, say, 0.2 wide;
for a galaxy evolution problem requiring the knowledge of the redshift
distribution on a much smaller scale redshift classes can be associated
with much smaller redshift bins.
Yet another ClassX classifier can use both types of redshift classes, 
treating galaxies and AGN as two sets of redshift classes.
Ultimately the class selection can be optimized so as to yield the
highest possible redshift resolution at an acceptable level of 
completeness and purity of redshift estimation.  In practice there 
is always, of course, a constraint that the number of exemplars of
each class is large enough to train the classifier. Because of that
we cannot have redshift classes with widths, say, of 0.0001.  

In this paper, redshift classes for galaxies are defined as 
redshift bins that are $\Delta z = 0.05$ wide. 
Because at high redshifts there are too few objects to split into 
smaller bins, the last bin is selected to cover a larger redshift range, 
$0.4 < z < 0.8$.  Similarly, each AGN redshift class is defined as 
a bin $\Delta z = 0.2$ wide, except that the last bin formally  covers
the range $z=2.6 - 6.0$.

For class attributes we select the SDSS photometric type 
(3 for resolved objects, 6 for point sources), and five
colors: $u-g$, $g-r$, $r-i$, $i-z$, and $g-i$; these are 
four main SDSS colors (see, e.g., Richards et al.\ 2004), and
the fifth color, $g-i$, is added to match the number of photometry
attributes to the number of independent  photometry bands. This
particular selection is not the only one possible, nor is it rigorously
justified, but our experiments showed that with these parameters
we obtain quite a good classifier.
One could have included more color indices or used magnitudes
instead of (or along with) colors. Previous experience in designing 
classifiers indicates, however, that having too many 
attributes, especially ones that are similar to each other or 
represent linear combinations of other attributes, increases 
the computational load while gaining little in terms 
of the classifier accuracy.

\subsection{Validation of ClassX Results} 

In order to validate the ClassX results,  we ran the software on the
validation sample described in \S\ref{training-and-validation}.  The
comparison of the ClassX results with the results from spectroscopy for
a set of 20,253 objects from that sample is given in Table~\ref{t_pref}.
The objects in that set are constrained to the magnitude range of the
bright photometric sample. At these magnitudes the number of galaxies 
beyond $z = 0.4$ is very small, so to make Table~\ref{t_pref}
 easier to read, we omitted the respective redshift classes in it.

The diagonal elements in Table~\ref{t_pref} give classification completeness,
i.e.,  the percentage of objects of a given class that ClassX identified 
as belonging to that class. ClassX correctly classified $\sim\!98.1$\% stars, 
$\sim\!98.5$\% galaxies, $\sim\!96.5$\% AGN, and 61.7\% red stars. 
M and later stars are frequently misclassified as
early-type stars (21.5\%) and intermediate redshift galaxies (12.2\%).
This is not surprising, because the number of M stars in the training set is
relatively small and there is a significant 
color overlap between the indicated classes.

 \subsection{Classifier Preference 
\label{s_completeness}}

The matrix  $\xi_{ij}$ given in Table~\ref{t_pref}, called 
{\it classifier preference} (McGlynn et al.\ 2004; Suchkov et al.\ 2003),
demonstrates how the classifier does class assignment.
If the classifier is given a sample of stars, its first  
preference for the sample objects  will be class
star, with  98.1\% of objects assigned to that class. 
The second preference will be class galaxy in the redshift range
$z=0 - 0.05$, with 0.5\% stars assigned to that class, 
and so on.
  
The matrix $\xi_{ij}$ tells us not only how good the classifier
is in distinguishing objects of a given class (diagonal elements) but
also  provides us with information as to where and in what numbers the
misclassified objects go (non-diagonal elements).  Therefore, it allows
us to analyze and quantify completeness and contamination of 
class populations derived by ClassX from photometric samples. 
For instance, we infer from Table~\ref{t_pref} that 0.5\% of stars 
in a sample under classification will be classified as low-redshift
galaxies, $z_{\rm{clx}} < 0.05$. If we have an estimate for the number of stars
in the sample, we can straightforwardly estimate the number of star 
contaminants among the objects classified as low-redshift galaxies.
From the training data we know exactly what kind of stars are 
misclassified as galaxies. Therefore, 
we would also know what kind of objects classified as low-redshift 
galaxies can be most easily confused with stars, which would allow us
to isolate such objects in the classification results and examine them
separately.

Classifier preference can be calculated as a function of
various parameters of interest and for various parameter constraints. 
As a result,  completeness of a ClassX classification can be 
quantified, for example, as a function of magnitude and/or other parameters;  
one can also account in an intelligent way for much of the contamination 
in the classification results using the contaminants properties 
derived from the training and validation samples.  

\subsection{ Redshifts from ClassX}  

\begin{figure}
\epsscale{0.5}
\plotone{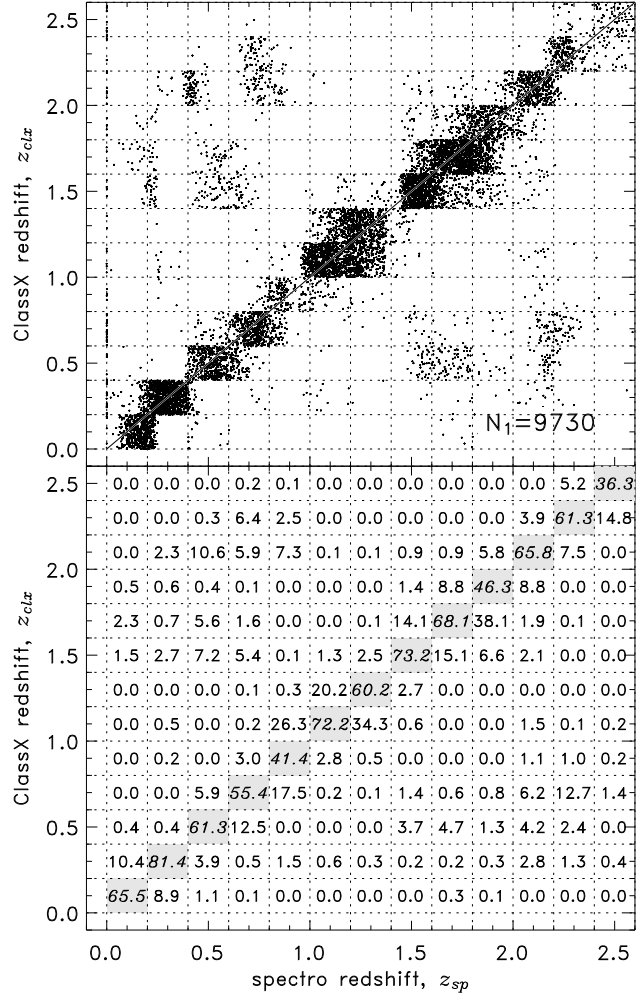}
\caption{ClassX redshifts, $z_{\rm{clx}}$, versus spectroscopic redshifts, 
$z_{\rm{sp}}$, for objects from the validation sample of 10,000 spectroscopic 
AGN (for better visualization, the ClassX redshifts are randomly distributed 
within each redshift bin and all redshifts larger than 2.4 are shown 
in the bin centered at 2.5). The lower panel displays 
the fraction (in percent) of AGN in a given spectroscopic redshift
bin assigned to different ClassX  redshift bins. 
The diagonal elements give the fraction of 
correctly classified redshifts, the non-diagonal elements give the fraction
of misclassified redshifts. Because a fraction of AGN is misclassified 
into stars and galaxies, the numbers in the columns in the lower panel
do not sum up to 100. Several clusters of points seen far from the diagonal
comprise less than 1\% of redshifts; they are due to 
``photometric degeneracy'', i.e., the regions in the color space
where the distinction between high and low redshift objects is small
(e.g., Richards et al. 2001b).   
\label{f1} }  \end{figure}

\begin{figure}
\epsscale{0.7}
\plotone{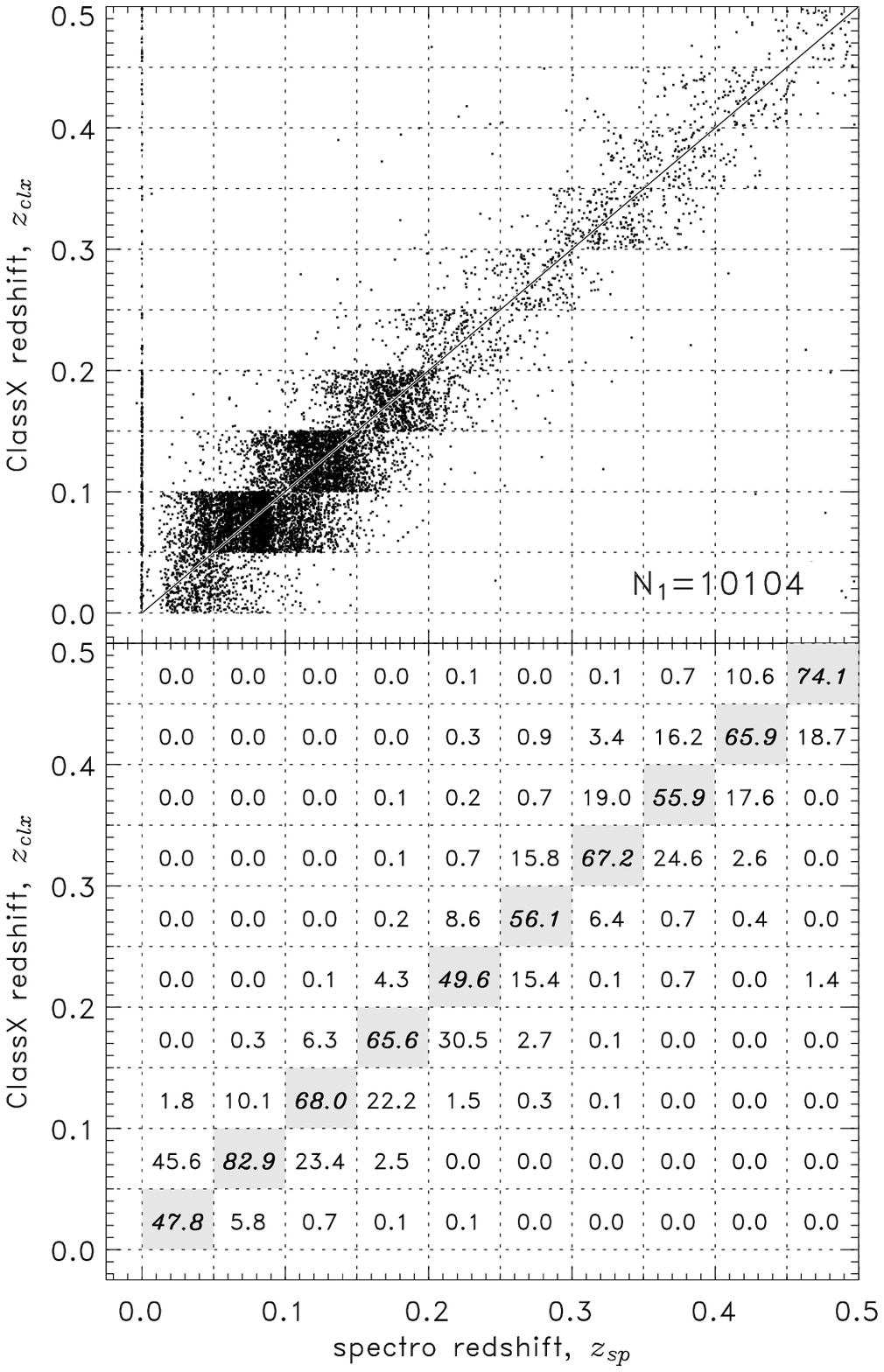}
\caption{Same as in Figure~\ref{f1} but for galaxies. 
\label{f2} }  \end{figure}

The SDSS photometric system was designed so as to allow one to get 
good redshift estimates directly from colors
(Richards et al.\ 2001a; 2001b; Budavari
2001). Richards et al.\ (2001a) demonstrated how the SDSS colors are 
influenced by redshift, and discussed in detail how individual 
features in AGN spectra contribute to the redshift information contained
in these colors. The clear understanding of the redshift effect in the SDSS 
colors allowed Richards et al.\ (2002) to construct
an efficient algorithm for selecting AGN candidates from the photometric 
database for subsequent spectroscopy. It was utilized in the SDSS DR2
and  resulted in $\sim\!3.6\times 10^4$ spectroscopically 
confirmed AGN (Abazajian et al.\ 2004a). 
Using a range of photometric 
redshift techniques, Csabai et al.\ (2003) determined
redshift estimates for six million SDSS EDR resolved objects (galaxies)
and gave an analysis of the statistical and systematic uncertainties.

In this paper we use redshifts of the SDSS DR2 spectroscopic AGN
and galaxies to train the ClassX classifier described above to
distinguish simultaneously both the  object type and its redshift
solely from the object colors and morphological type.
Figures~\ref{f1}~and~\ref{f2} show how well the 
classifier discriminates redshifts of photometric objects 
classified as AGN and galaxies.  One can see that
most of the ClassX redshifts are in the same bins 
(classes) as the redshifts from spectra, i.e., on the diagonal, 
and the misidentified  redshifts appear
mostly in the bins adjacent to the diagonal bins. Overall  
the classifier estimates redshifts reasonably well
at the adopted level of redshift resolution and compares
quite well with other redshift photometric techniques
(see Table~\ref{tnew}). Of course, as a generic classifier, 
ClassX does much more than just estimating redshifts. 

Similar to other photometric methods, the classifier performance 
in redshift estimation is uneven across the redshift bins. But
analysis of the results easily provides ideas as to how the classifier
design can be changed to improve on that. For instance, of all AGN with 
true redshifts in the range $0.8 < z_{\rm{sp}} < 1.0$, only 41.4\% were 
assigned correct redshifts while 43.8\% of the remaining AGN  were placed 
into the two adjacent redshift bins; this is obviously below the average 
success rate for this classifier (see Table~\ref{tnew}).
One can notice, however, that the bulk of the misclassification 
into the bin $1.0 < z_{\rm{clx}} < 1.2$ occurs from a narrow range, 
$\sim 0.95 < z_{\rm{sp}} < 1.0$ (see Figure~\ref{f1}). Similarly, 
misclassification into the bin $0.6 < z_{\rm{clx}} < 0.8$ occurs
mostly from the range $\sim 0.80 < z_{\rm{sp}}< 0.90$. This suggests 
splitting of the redshift class $0.8-1.0$ into two new redshift
classes divided at $z = 0.9$ may substantially improve the
classifier redshift resolution. 

Figures~\ref{f1}~and~\ref{f2} allow one to examine in
detail how the color information used by ClassX becomes 
ambiguous for certain redshifts. Similar to  the
above discussion, we see that the AGN redshifts are misplaced 
from the $0.2-0.4$ range to the $0-0.2$ range, but not in 
a uniform way.  Rather, only the ones within the narrow range 
of $0.20 \leq z_{\rm{sp}} \la 0.25$  are confused for redshifts
in the $0 - 0.2$ range; there is a substantial confusion of redshifts
in the  range $z_{\rm{sp}} \sim 0.40 - 0.45$ with redshifts 
$z_{\rm{clx}} = 2.0 - 2.2$, and so on.  Comparing Figure~\ref{f1} with 
similar diagrams in Richards et al.\ (2001b; 2004) one can notice many 
common features in the area where the diagrams overlap, which indicates 
that the ClassX redshift misidentifications reflect the same photometric
degeneracy that  was discussed by Richards et al.\ (2001b).
This degeneracy is inherent to the data; however, its impact on
a classifier can be reduced to a minimum by a careful selection 
of classes in general and a careful design of redshift classes
in particular.  In general, if redshift resolution
and accuracy is the goal, a much more sophisticated classifier
can be designed that will meet that goal.

\begin{figure}
\epsscale{1.0}
\plotone{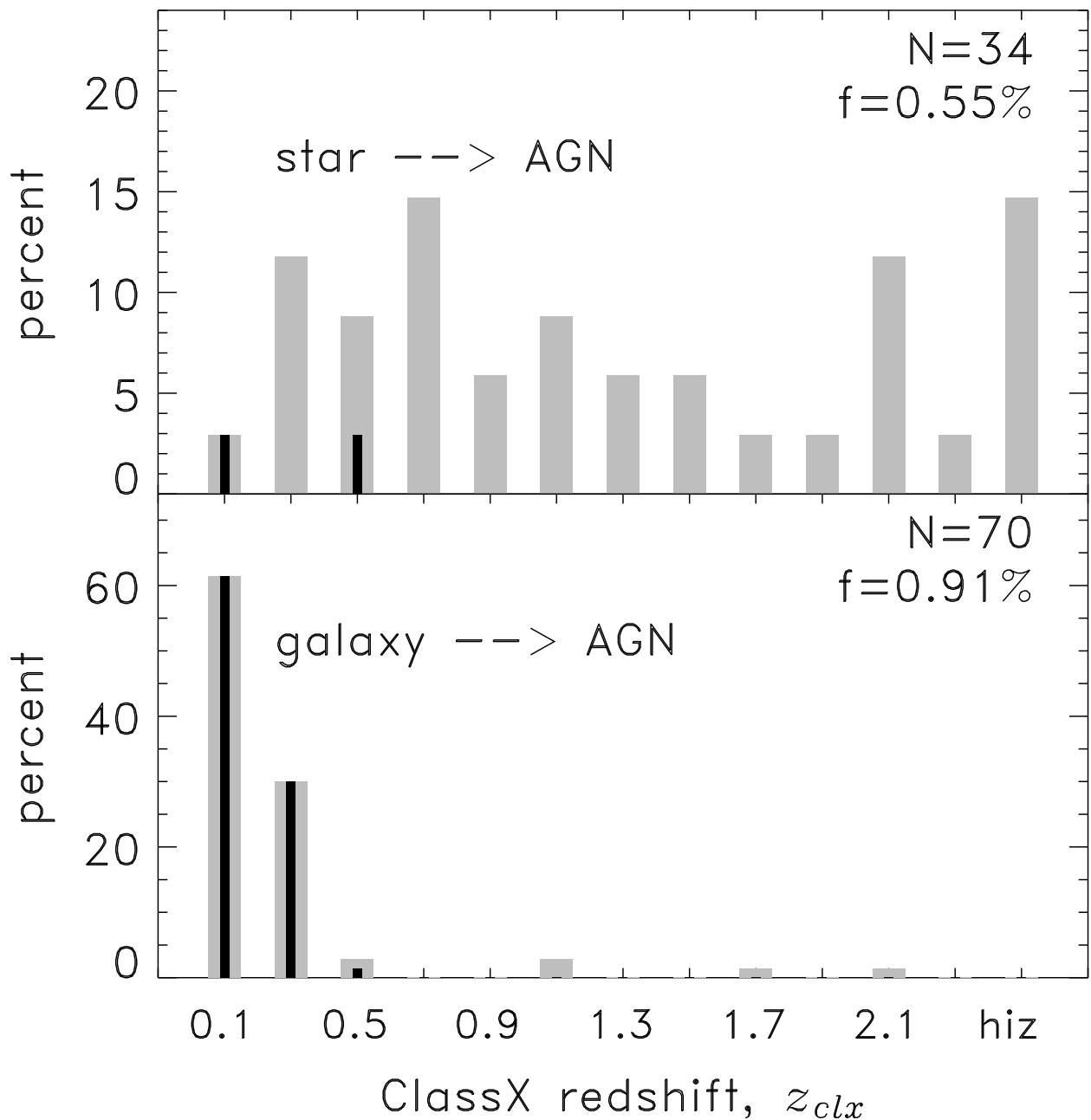}
\caption{
Distribution of redshifts assigned by ClassX to spectroscopic 
stars and galaxies 
misclassified as AGN. Both the stars and the galaxies are 
from the respective subsamples of spectroscopic objects not used in 
the classifier training 
and constrained to  the magnitude range of the bright photometric sample;  
$f$ is the percentage of the misclassified stars or galaxies 
in the subsample. The central black bar indicates the contribution of
resolved objects. Only a tiny fraction of stars and normal galaxies is 
misclassified as AGN. Also note that misclassified stars  
are scattered across the entire range of AGN redshifts, while 
almost all misclassified galaxies get only low redshifts, 
$z < 0.4$. 
\label{f3} } \end{figure}

\begin{figure}
\epsscale{1.0}
\plotone{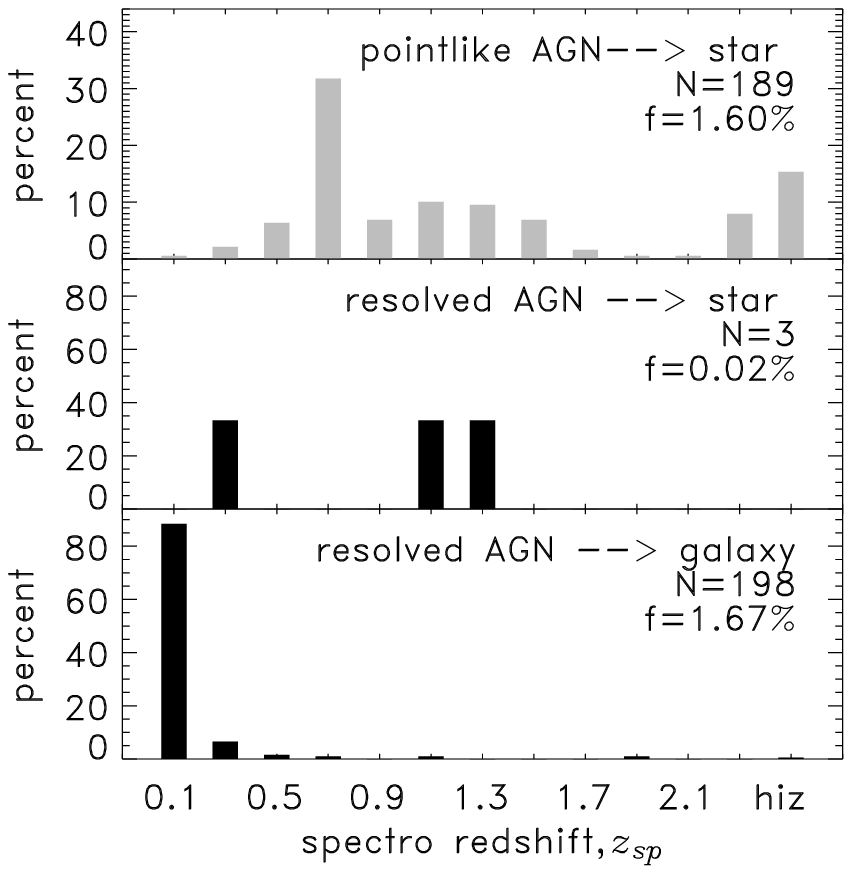}
\caption{
Redshifts of AGN misclassified  as stars and galaxies.  The AGN are 
from the spectroscopic sample constrained to  the magnitude range 
of the bright photometric sample; $f$ is the percentage of the 
AGN misclassified into the respective class. None of the pointlike AGN
was classified into class galaxy.  Note that most of the misclassification
of resolved AGN occurs from the lowest redshifts, $z_{\rm{sp}} < 0.2$. 
\label{f4} } \end{figure}

We infer from Table~\ref{t_pref} that some small fraction of 
AGN from ClassX would be  misclassified as stars and galaxies.
What redshifts does ClassX  assign to these contaminants?
Figure~\ref{f3} gives the answer. While the stars 
are scattered across the entire range of AGN redshifts, 
with a marginal preference for the high-redshift 
bin, ``hiz'', and the $z_{\rm{clx}} = 0.6 - 0.8$ bin, the misclassified 
galaxies are placed by the classifier mostly into the
low-redshift range, with 60\% of galaxy contaminants
having $z_{\rm{clx}} < 0.2$. So if we see that AGN from ClassX
are more numerous within $z_{\rm{clx}} = 0.2 - 0.4$ than at $z_{\rm{clx}} < 0.2$,
we would know that this is not the effect of galaxy contaminants. 

The redshift distribution of AGN misclassified into stars and galaxies
is shown in Figure~\ref{f4}. Not surprisingly, misclassification
of pointlike AGN occur into class star, while misclassified resolved AGN 
turn out to have the lowest redshifts and end up in class galaxy.
This means that almost all incompleteness in classified resolved AGN
will be at low redshifts. Incompleteness in classified pointlike AGN,
on the other hand, will be due mostly to objects within the range
$z_{\rm{sp}} = 0.6 - 0.8$.
 
\section{Population Content of the SDSS Photometric Database}

\subsection{SDSS Major Spectroscopic Object Classes in the Photometric Catalog}

Table~\ref{t_content} summarizes the results from 
classification of the three photometric samples, indicating  
the fraction of different object types in the 
SDSS photometric data base and how that fraction varies as 
a function of magnitude constraints. Stars dominate all three
samples. Their fraction substantially decreases toward fainter
magnitudes, dropping to $ 80\%$  in the faint sample. The fraction
of red stars, however, increases rather than decreases as the magnitudes
get fainter, and in the faint sample it is nearly twice as large as
in the bright sample.     

Comparing the bright and intermediate 
magnitude limits, we see that going 1~mag fainter in all bands
increases the fraction of AGN by a factor of 3, from 1.06\%
to 3.16\%. The fraction of galaxies experiences a much less
dramatic variation, increasing only by a factor of 1.4. 
Moving to the faint brightness range, we see that the fraction of 
AGN increases substantially again, by more than a factor of 2
in comparison with the intermediate magnitude range. At the same time 
the fraction of galaxies changes very little.   

In the bright magnitude range, the ClassX estimated number of AGN 
in the SDSS DR2 photometric catalog is $\sim\! 4.0 \times 10^4$
(Table~\ref{t_content}). 
This compares well with the number of AGN in the 
SDSS DR2 spectroscopic catalog, $\sim\! 3.6 \times 10^4$,
especially if one takes into account the fact that the sky coverage
of the spectroscopic catalog is a bit smaller, 2627~deg$^2$
versus 3324~deg$^2$ for the photometric catalog. Toward
fainter magnitudes, the number of AGN goes up dramatically.
It increases to $\sim\! 2.0 \times 10^5$ as we move over 
to the intermediate  magnitude range, and becomes
$\sim\! 5.0 \times 10^5$ in the magnitude range of the faint sample.

These estimates refer only to objects
with clean photometry as defined by the selection criteria
for the bright, intermediate, and faint photometric samples.  
Also, the statistics 
derived for our samples are driven by the adopted set of magnitude 
constraints, which directly impact the derived number counts and completeness
of class objects (see also \S\ref{sec-samples}). Lowering the magnitude 
limit for the $u$~band while
keeping the other limits the same would increase the fraction of AGN objects 
in the sample, because this would include more sources with $UV$~colors 
typical for AGN and rare among stars and galaxies. Similarly, 
constraining a sample to brighter magnitudes in the red bands  at the same
magnitude limits in the $UV$ and blue bands would result in a larger
fraction of red galaxies and red stars. These examples 
illustrate that before interpreting classification results 
in terms of actual physics one has to properly analyze and take into 
account the sample selection effects.

\begin{figure}
\epsscale{1.0}
\plotone{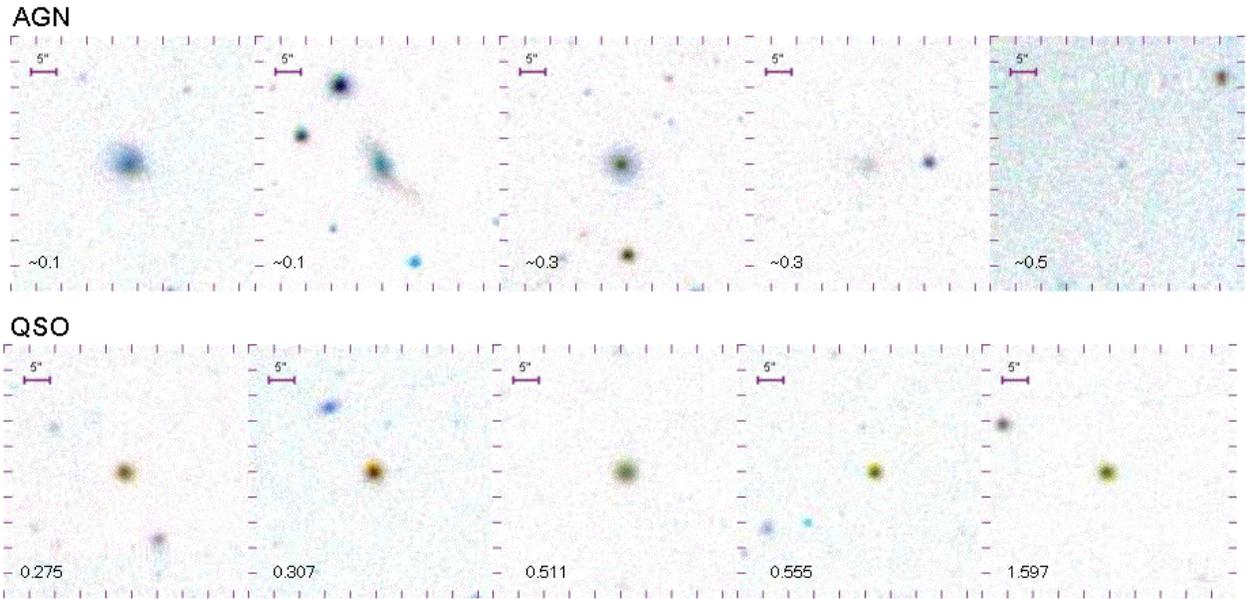} 
\caption{
A comparison of SDSS images of typical resolved AGN (top) and pointlike
AGN (bottom) as classified by ClassX. 
The objects are from the magnitude range $18 < g < 20$. 
For unresolved AGN, the spectroscopic redshift is shown (these objects were 
also found in the spectroscopic catalog), and for the resolved AGN, only 
the ClassX-estimated redshift is 
given (none of these AGN candidates were found in the 
spectroscopic catalog).  The resolved AGN show 
irregular structure and concentrated nuclear emission.
\label{f5} } \end{figure}

It is useful to compare the AGN surface density based on 
the results from ClassX with previously known similar estimates 
for SDSS objects; this allows us to make a consistency check
across different methods of estimation of AGN number counts.
Richards et al.\ (2002) used a sample of known AGN brighter
than $i \sim 19$  to determine the AGN sky density,
which they found to be $\sim\! 14$~deg$^{-2}$. 
This number is consistent with the results of the 
AGN spectroscopic survey for the SDSS DR2, 
$S_{\rm{sp}} =  3.6 \times 10^4/2627 \approx 13.7$~deg$^{-2}$
(which implies a high efficiency of the Richards et al.\ 2002 
AGN selection algorithm).
Using the ClassX object number estimates in Table~\ref{t_content},
we can calculate the  AGN surface density for the magnitude constraints
given in Table~\ref{t_photo_samples}.
For the magnitude ranges of the bright, 
intermediate, and faint photometric samples
 we get $S_{\rm{bright}} = 12.2$~deg$^{-2}$,
$S_{\rm{intermediate}} = 58.4$~deg$^{-2}$, and 
$S_{\rm{faint}} = 150.3$~deg$^{-2}$.

SDSS provides a parameter, morphological type, that distinguishes 
resolved (extended) and unresolved (pointlike) sources in SDSS images.  
It is useful, in particular, for isolating AGN objects that clearly
exhibit the extended component of the underlying galaxy.
The morphological differences between resolved and unresolved AGN
from ClassX classification  are illustrated in Figure~\ref{f5}. 
With resolved and unresolved AGN separated in the classification
results, we can compare the statistics of the two morphological types
with the results from other studies.  For example, our faint sample 
has the same $g$-magnitude limit as the catalog of QSO (point-source) 
candidates derived by Richards et al.\ (2004) from the SDSS DR1 
photometric catalog. The QSO surface density  estimated in that paper 
is 45~deg$^{-2}$. This value compares quite well with 
$S_{\rm{faint}}({\rm{AGN_{\rm unresolved}}}) = 52.6$~deg$^{-2}$ 
that we obtain for the faint sample. 

Richards et al.\ (2004) noted that the surface density of quasars 
in their catalog is substantially larger than the density of similar
objects from the Schneider et al.\ (2003) catalog of spectroscopically 
identified quasars, 45.5 versus 6.95~deg$^{-2}$. They concluded 
that their large QSO sample will, therefore, be very powerful for 
investigations of problems such as quasar-quasar and quasar-galaxy
clustering. It is obvious that similar conclusions apply to the AGN 
samples from a classification of SDSS DR2 with ClassX. 
For instance, 
one can investigate clustering of normal and AGN 
galaxies along the lines similar to the recent paper by
Zehavi et al.\ (2005). Given that the ClassX samples 
are robustly constrained in terms of limiting magnitudes and are well 
quantified in terms of completeness, they are exceptionally well
suited for analysis of such fundamental problems as
the evolution of the AGN luminosity function.

\begin{figure}
\epsscale{1.0}
\plotone{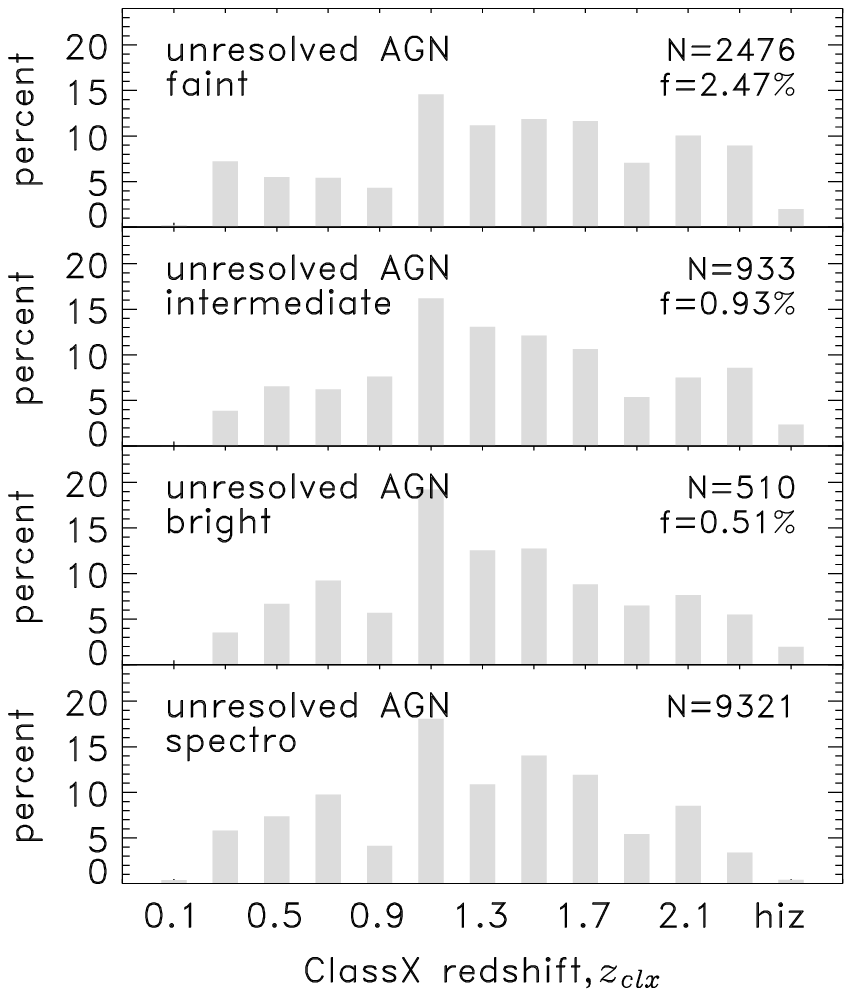} 
\caption{
Distribution of ClassX  redshifts for pointlike AGN. $N$ is the number
of objects classified as AGN in the given sample, and $f$ is
the fraction of these objects, $f=N/N_{\rm sample}$. 
The  bright and intermediate samples 
exhibit a noticeable  downtrend after the peak at $z_{\rm{clx}}\sim 1.0-1.2$;
the faint sample shows a more even distribution  in this range. 
\label{f6} } \end{figure}

\begin{figure}
\epsscale{1.0}
\plotone{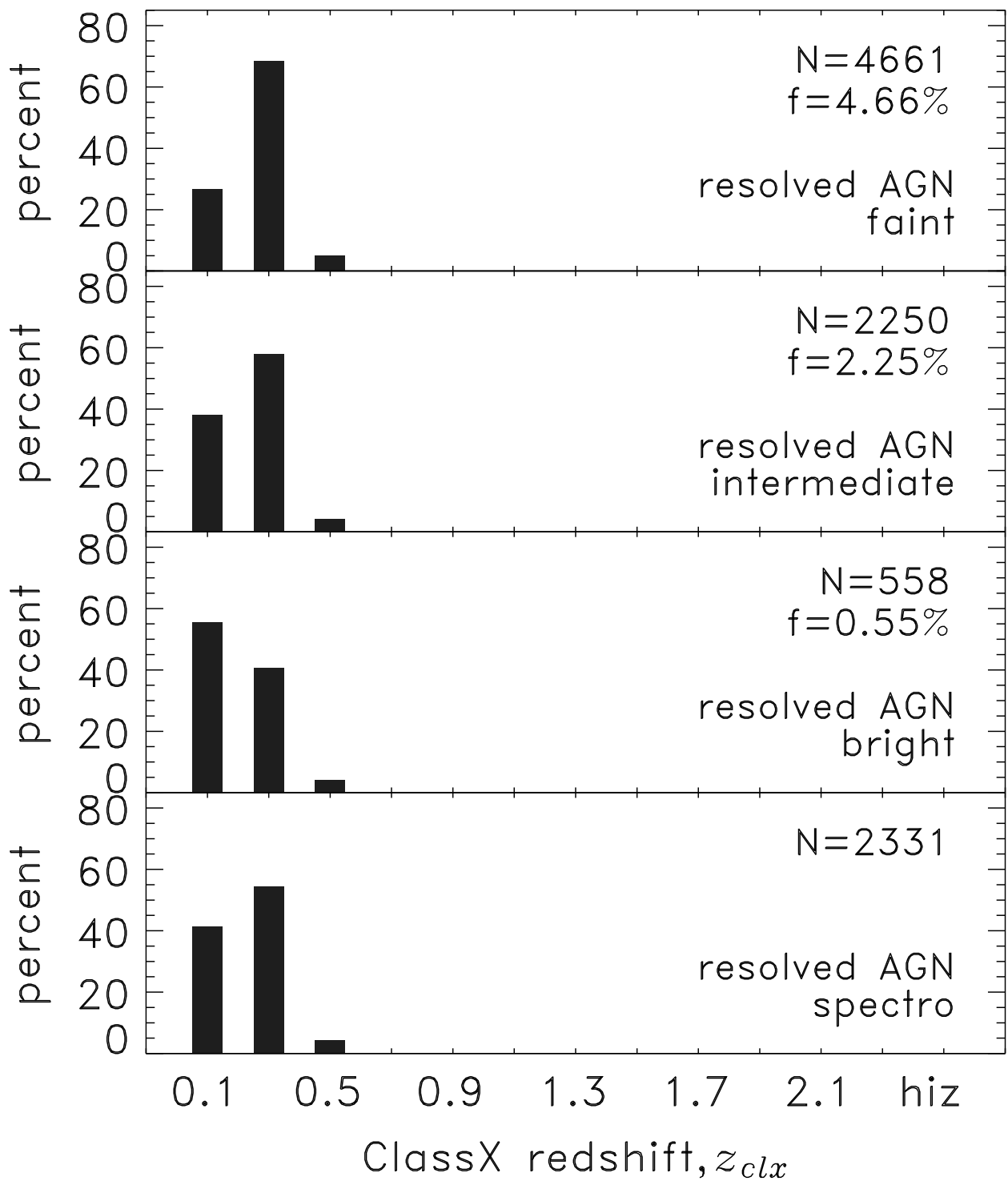} 
\caption{
Same as in Figure~\ref{f6} but for resolved AGN. 
Toward  fainter magnitudes (intermediate and faint samples),  a large
number of AGN appears in the redshift range 0.2--0.4. 
\label{f7} } \end{figure}

\begin{figure}
\epsscale{1.0}
\plotone{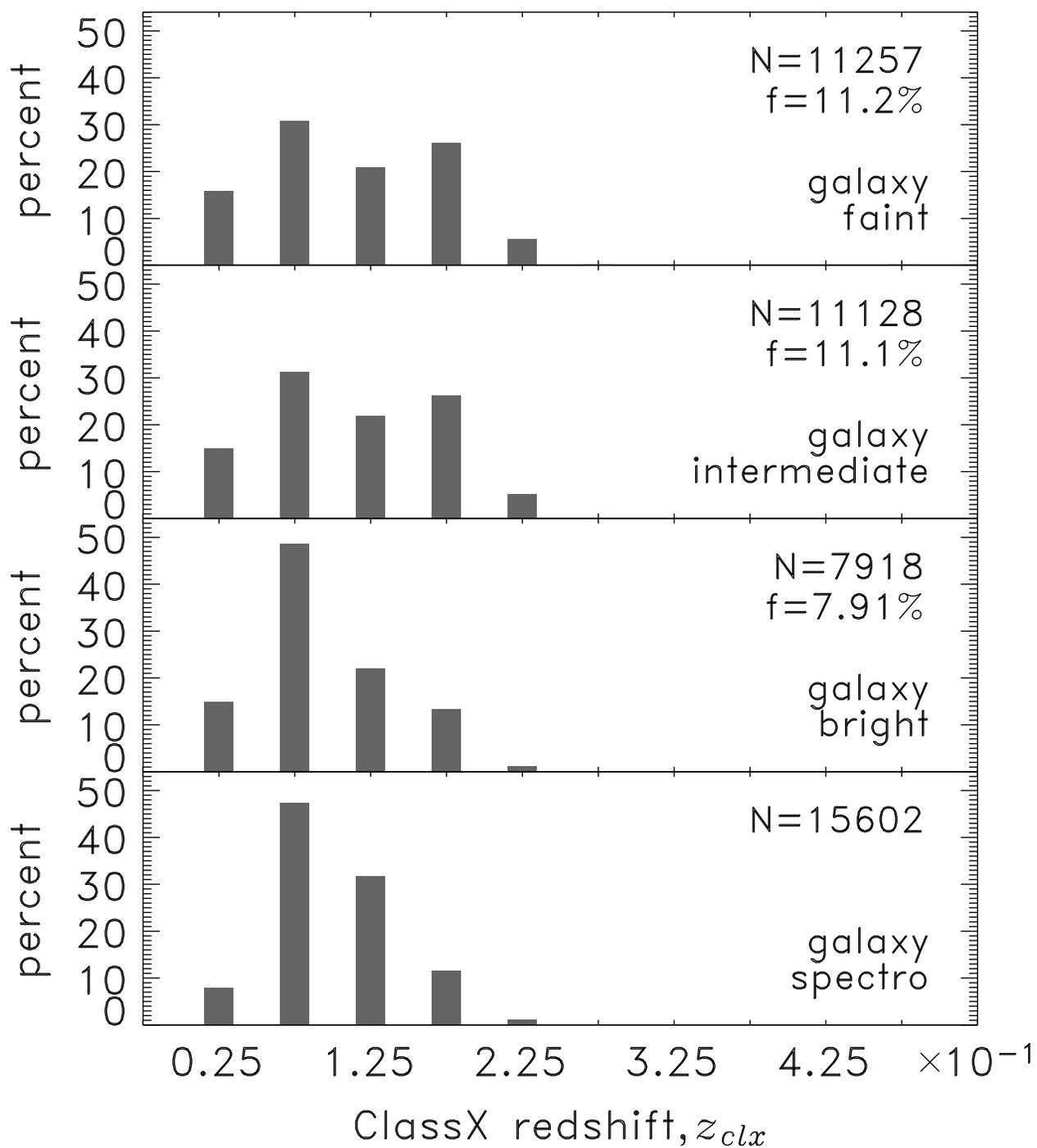} 
\caption{
Same as in Figure~\ref{f6} but for galaxies.
At faint magnitudes, more galaxies appear at higher redshifts.
Yet AGN outnumber them at redshifts $z_{\rm{clx}} > 0.2$ 
in the  brightness range of the intermediate and faint samples.  
\label{f8} } \end{figure}

\subsection{Redshift Distribution of Galaxies and AGN as a Function
of Magnitude Constraints}

There is a substantial redshift dependence in the variation of the
number of AGN and galaxies as a function of magnitude constraints (see 
Figures~\ref{f6}--\ref{f8}).  In the bright magnitude range,
the fraction of the resolved AGN within the redshift range $0.2 < z < 0.4$
is $40$\% of the resolved AGN population. Such AGN become, however, 
dominant in the intermediate and faint samples, in which their 
fraction jumps to $60$\% and $70$\%, respectively.

Due to the selection effects caused by magnitude constraints,
resolved AGN and normal galaxies are represented very differently in
our photometric samples. Thus, resolved AGN are more numerous at redshifts
$z > 0.2$, because non-AGN galaxies are intrinsically too faint in the UV
and the bulk of them have the $u$-brightness below the limiting magnitude
in the $u$-band.  Only $\sim\!1$\% of objects that we classified as
normal galaxies in the bright sample have redshifts $z > 0.2$. 
This fraction rises to $5$\% in the faint sample, yet the 
number and surface density of resolved AGN in that magnitude range 
is $5$~times larger than that of galaxies. 
Figures~\ref{f7}~and~\ref{f8} illustrate the difference in the rate of increase
in the number of galaxies and resolved AGN toward the faint sample. 
While the variation in the number of galaxies between 
the bright and faint samples is less than a factor of 1.5. 
the number of resolved AGN increases by a much larger factor of 8.3.   

AGN can be observed as resolved objects in SDSS only at relatively 
low redshifts.
Unlike pointlike AGN, which are distributed more or less evenly over 
the redshift range extending up to $z \sim 2$, they are almost entirely 
confined to redshifts $z < 0.4$ (Figures~\ref{f6}~and~\ref{f7}). 
More interesting is the fact that the number of resolved AGN increases 
toward the faint sample considerably faster than the number of pointlike AGN. 
Between the bright and faint samples, the number counts of resolved AGN 
increases by  a factor of 8.3 while the increase in the number counts 
of pointlike AGN is less than a factor of 4.8. It is tempting to try 
to interpret this effect in terms of cosmological evolution of AGN
and/or AGN--galaxy connections. However, before doing that, 
one has first to establish the biases caused by the sample magnitude
constraints (this is true, of course, 
with respect to essentially all statistical properties seen 
in the immediate results from ClassX classification and redshift estimation).
We intend to conduct such an analysis in a follow-up paper.  

\section{Resolved SDSS AGN and Galaxies at Faint Magnitudes}

\begin{figure}
\epsscale{1.0}
\plotone{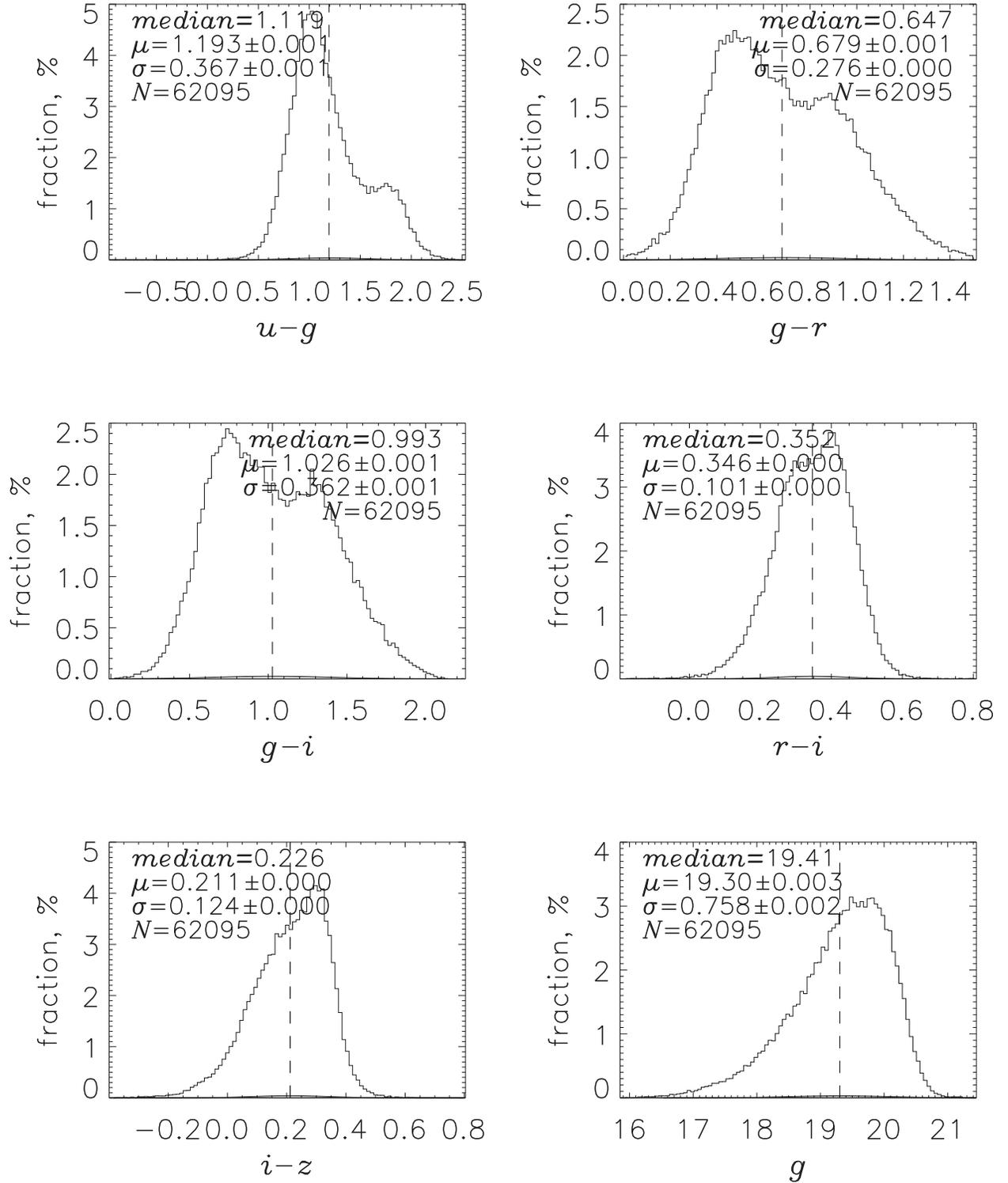} 
\caption{
Color distributions of candidate normal galaxies from Table~\ref{t-catalog}.
The distributions show characteristic features of color 
distribution of SDSS galaxies, such as bimodality of the $u-g$ distribution. 
Also shown is the brightness distribution in the  $g$-band, which is to be 
compared with the similar distribution of candidate AGN galaxies in 
Figure~\ref{f10}.
(Because of truncation of the color ranges,
the number of objects in each plot is somewhat smaller than in 
Table~\ref{t-catalog}). The legend gives the median, mean, and standard
deviation for each distribution. 
\label{f9} } \end{figure}

\begin{figure}
\epsscale{1.0}
\plotone{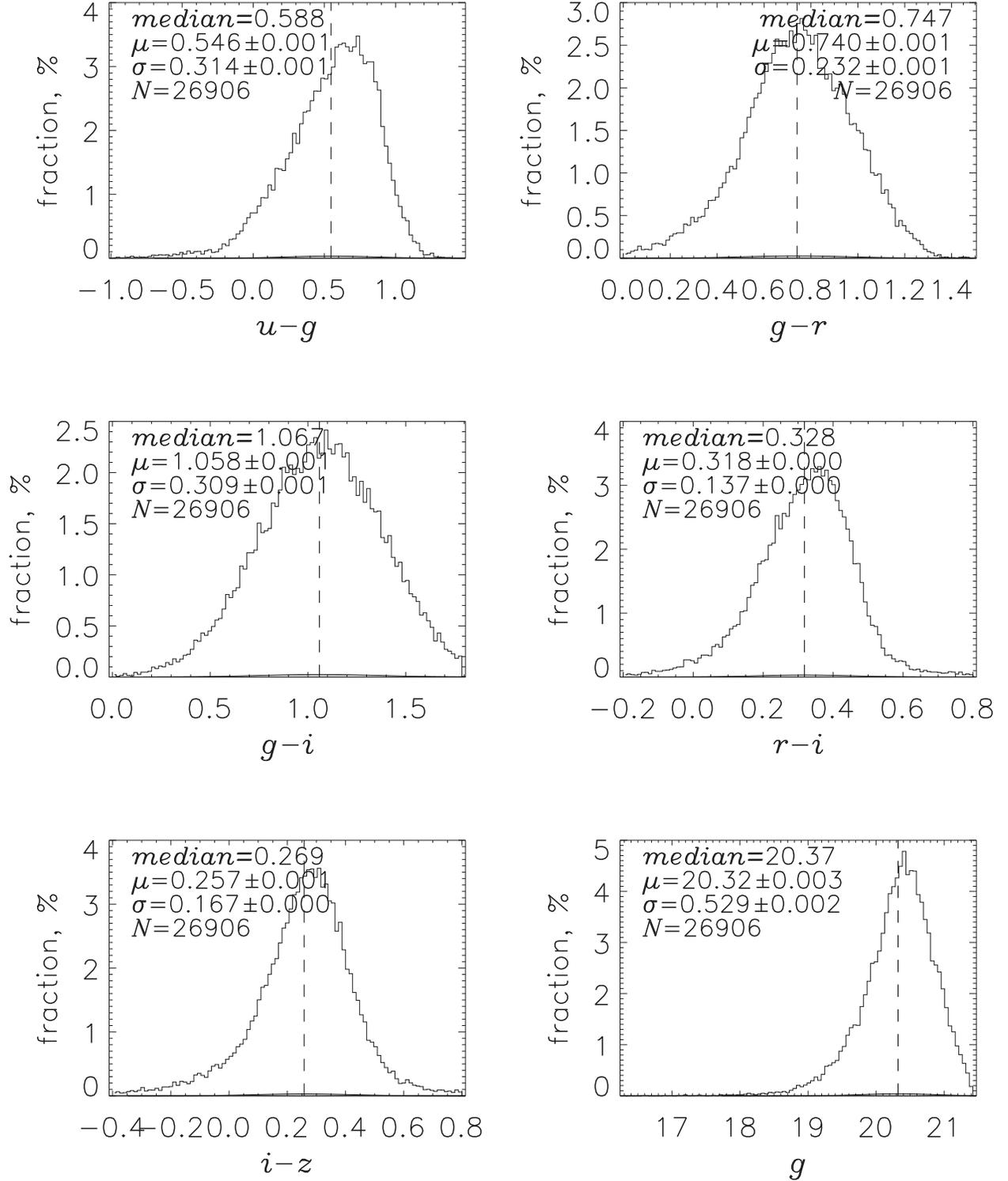} 
\caption{
Same as in Figure~\ref{f9} but for resolved AGN.
Unlike the case of normal galaxies, the AGN color distributions
show no bimodality.  As expected, the objects are on average
 much bluer in $u-g$, being at the same time noticeably redder 
$g-r$. The differences in brightness distributions are illustrated 
by the magnitude distribution in the $g$-band. The 
AGN candidates are on average fainter than the normal 
galaxy candidates by $\sim\! 1$~mag in $g$, which is also expected for
the magnitude constraints imposed on the sample.
\label{f10} } \end{figure}

As an illustration of ClassX application to SDSS data,
we provide in Table~\ref{t-catalog} a sample catalog of $91,847$ 
resolved objects from the SDSS DR2 photometric data base 
classified by ClassX into normal and AGN galaxies.
While there are nearly 265,000 galaxies in the SDSS DR2 spectroscopic
catalog, only $\sim$4,000 objects in that catalog are 
spectroscopically identified resolved AGN. 
Our catalog in Table~\ref{t-catalog} contains  29,005 candidate 
AGN resolved in SDSS imaging, seven times more than in the DR2 spectroscopic
catalog. Thus, ClassX can easily produce a huge new resource for studying 
AGN galaxies and their relationships with normal galaxies.

The objects in Table~\ref{t-catalog} are from a
representative sample of 100,000 resolved objects from the SDSS DR2
photometric catalog. The magnitude and data quality constraints are the
same as for the faint sample in Table~\ref{t_photo_samples}. There are 
1,482,310 resolved objects in the SDSS DR2 photometric catalog satisfying 
these constraints, so our sample comprises $6.75$\%  of them.  
Along with 29,005 of the sample objects classified as AGN, 
62,842 objects were classified as galaxies; $8\%$ of objects were 
assigned class star or red star and were not included into
Table~\ref{t-catalog}. 

The percentage of candidate normal galaxies and resolved AGN in
Table~\ref{t-catalog} is 68.4\% and 31.6\%, respectively. This compares
well with  70.7\% and 29.3\% for the faint sample
(see the numbers in Figures~\ref{f7}~and~\ref{f8}), supporting
the inference that about a third of all SDSS galaxies within the
magnitude regime of the faint sample harbor active nuclei. 

Normal galaxies and resolved AGN constrained in the same way
with respect to SDSS magnitudes are
very different in terms of their color properties. 
Figures~\ref{f9}~and~\ref{f10} demonstrate that
candidate normal and AGN galaxies in Table~\ref{t-catalog} 
exhibit such differences too. Normal galaxies show a pronounced  
bimodality of color distribution 
in $u-g$, $g-r$, and $g-i$, while nothing of that is observed
in the color distributions of candidate AGN galaxies. Resolved
AGN are, as expected, substantially bluer in $u-g$ and
noticeably  redder in $g-r$. In the $g$~band, they are 
on average fainter than normal galaxies by $\sim\! 1$~mag, which is also
consistent with the expectation from the magnitude constraints.

It is to be noted that the large fraction of AGN objects in 
Table~\ref{t-catalog} is due to the specific magnitude constraints
imposed on the sample, and the effect of constraints varies with
redshift. In particular, our magnitude limits strongly  
favor AGN galaxies with redshifts $z > 0.2$ and work
against normal galaxies at these redshifts. This explains why candidate
active galaxies in Table~\ref{t-catalog} outnumber normal galaxies at
these redshifts by a ratio of more than 5 to 1 (21,146 
resolved AGN candidate versus 3,967 normal galaxies).

\section{Conclusions and Future Work}

ClassX provides an effective classification of SDSS photometric objects 
into stars, galaxies, and AGN, and yields quite accurate redshifts 
for the bulk of galaxies and AGN. When tested on a sample
of $\sim\!20,000$ spectroscopically identified objects from the SDSS DR2 
spectroscopic catalog, it correctly classified 98.1\% of the stars, 
98.5\% of the galaxies, and 96.5\% of the AGN. The ClassX approach is 
applicable to any object class with sufficient representation in the 
SDSS, and thus complements class-specific selection algorithms such as
used by Richards et al.\ (2004).

 We classified a set of representative samples from 
the SDSS DR2 photometric catalog and obtained estimates of the catalog 
population content in different magnitude ranges. We used redshifts 
from ClassX to compare redshift distributions of the catalog objects  
classified as galaxies and AGN. As an illustration of ClassX applications,
we provide a sample catalog of resolved objects from the SDSS photometric
catalog that contains 27,000 candidate AGN galaxies along with 
63,000 objects classified as normal galaxies.

Future work will include both the creation of more powerful ClassX
classifiers and interpretation of the classification results.  
With newer releases of SDSS data, we can use much larger training sets
and expand the class sets handled by classifiers. Redshift estimates
will be refined by exploiting  a more intelligent selection of redshift 
classes based on the results of the present study. Inclusion and analysis 
of more object types  should help to interpret the 
classification results. For instance, there is an obvious need to  
incorporate starburst galaxies to disentangle a potential confusion
of the AGN and starburst phenomena (which is especially challenging
because the two often go together).  There is rich information
on resolved objects in the SDSS photometric database, such as morphology
parameters characterizing object light distribution in different bands.
Including these parameters in the sample attributes may help to 
discriminate normal galaxies, AGN, and starburst galaxies at magnitudes
beyond the  limit of the SDSS spectroscopic objects. 

\acknowledgements{
Funding for the creation and distribution of the SDSS Archive has been
provided by the Alfred P. Sloan Foundation, the Participating
Institutions, the National Aeronautics and Space Administration, the
National Science Foundation, the U.S. Department of Energy, the
Japanese Monbukagakusho, and the Max Planck Society. The SDSS Web site
is http://www.sdss.org/.  The SDSS is managed by the Astrophysical
Research Consortium (ARC) for the Participating Institutions. The
Participating Institutions are The University of Chicago, Fermilab, the
Institute for Advanced Study, the Japan Participation Group, The Johns
Hopkins University, the Korean Scientist Group, Los Alamos National
Laboratory, the Max-Planck-Institute for Astronomy (MPIA), the
Max-Planck-Institute for Astrophysics (MPA), New Mexico State
University, University of Pittsburgh, University of Portsmouth,
Princeton University, the United States Naval Observatory, and the
University of Washington.

Partial support for this work was provided by NASA's Applied Information
Systems Research Program (AISRP) under grant NAG5-11019 to the Universities
Space Research Association, subgrant 05095-01 to the Space Telescope Science
Institute.

We wish to thank T. Heckman for many comments and suggestions 
regarding the analysis of the ClassX extragalactic populations. 
We are grateful to G. Richards for reading the paper, revision 
suggestions, and valuable comments on SDSS morphological classification
at faint magnitudes.  
We are also thankful to I. Baldry, A.
Conti, and C. Leitherer for useful discussions of a possible role of
starburst galaxies; we thank T. Budavari for his comments on the paper
and sharing his insight into photometric redshifts from Sloan data.
Finally, we thank the anonymous referee for numerous comments and
suggestions that helped to improve the paper.
 }

\begin{deluxetable}{lccccc}
\tabletypesize{\small}
\tablewidth{0pt}
\tablecaption{\sc Photometric Samples 
\label{t_photo_samples}
}
\tablehead{
\colhead{}
& \multicolumn{5}{c}{Magnitude Range}
\\
\colhead{Sample\ \ \ \ \ \ \ }
& \multicolumn{5}{r}{\hrulefill}
\\
{} & $u$ & $g$ & $r$ & $i$ & $z$
\\
}
\startdata
Bright\dotfill
&17.0--20.5 & 16.0--19.5 & 15.5--19.0 & 15.0--19.0 & 14.5--19.0 
\\
Intermediate\dotfill
&17.0--21.5 & 16.0--20.5 & 15.5--20.0 & 15.0--20.0 & 14.5--20.0
\\
Faint\dotfill
&17.0--21.5 & 16.0--21.5 & 15.5--21.0 & 15.0--21.0 & 14.5--21.0
\\
\enddata
\end{deluxetable}

\begin{deluxetable}{l@{}r@{}@{}r@{}@{}r@{}@{}r@{}@{}r@{}@{}r@{}@{}r@{}@{}r@{}@{}r@{}@{}r@{}@{}r@{}@{}r@{}@{}r@{}@{}r@{}@{}r@{}@{}r@{}@{}r@{}@{}r@{}@{}r@{}@{}r@{}@{}r@{}}

\tabletypesize{\scriptsize}
\rotate
\tablewidth{0pt}
\tablecaption{{\sc Classifier Preference, $\xi_{ij}$, and Validation of ClassX Classification\label{t_pref} }}
\tablehead{
ClassX Class &  \multicolumn{20}{c}{True Class} \\
{} &  \multicolumn{21}{r}{\hrulefill} \\
\colhead{} & \colhead{01}  &  \colhead{02}  &  \colhead{03}  &  \colhead{04}  & 
\colhead{05}  &  \colhead{06}  &  \colhead{07}  &  \colhead{08}  & 
\colhead{09}  &  \colhead{10}  &  \colhead{11}  &  \colhead{12}  & 
\colhead{13}  &  \colhead{14}  &  \colhead{15}  &  \colhead{16}  & 
\colhead{17}  &  \colhead{18}  &  \colhead{19}  &  \colhead{20}  & 
\colhead{21}    
}
\startdata
01--star\dotfill & 98.1 &  3.3 &  0.4 &  0.1 &  0.5 &  0.7 &  0.0 &    0.2 &  0.5 &  1.1 &  7.4 &  1.3 &  1.9 &  1.8 &  1.0 &  0.2 &  0.0 &  0.3 &  4.7 & 47.2 & 21.5 \\
02--0.025\dotfill &  0.5 & 47.6 &  6.1 &  0.8 &  0.0 &  0.7 &  0.0 &   0.0 &  0.0 &  0.0 &  0.2 &  0.0 &  0.0 &  0.0 &  0.0 &  0.0 &  0.3 &  0.0 &  0.0 &  0.0 &  0.2 \\
03--0.075\dotfill &  0.4 & 47.8 & 82.7 & 25.9 &  3.7 &  0.0 &  0.0 &   2.7 &  0.0 &  0.4 &  0.0 &  0.0 &  0.0 &  0.0 &  0.0 &  0.0 &  0.0 &  0.0 &  0.0 &  0.0 &  4.4 \\
04--0.125\dotfill &  0.3 &  1.0 & 10.3 & 66.5 & 27.2 &  4.6 &  4.5 &   5.5 &  0.1 &  0.0 &  0.0 &  0.0 &  0.0 &  0.0 &  0.0 &  0.0 &  0.0 &  0.0 &  0.0 &  0.0 &  4.4 \\
05--0.175\dotfill &  0.1 &  0.0 &  0.3 &  6.2 & 62.1 & 56.2 & 13.6 &   7.5 &  0.5 &  0.0 &  0.0 &  0.0 &  0.0 &  0.0 &  0.0 &  0.0 &  0.0 &  0.0 &  0.0 &  0.0 &  3.4 \\
06--0.225\dotfill &  0.0 &  0.0 &  0.0 &  0.1 &  3.6 & 28.1 & 18.2 &   0.5 &  0.1 &  0.0 &  0.0 &  0.0 &  0.0 &  0.0 &  0.0 &  0.0 &  0.0 &  0.0 &  0.0 &  0.0 &  0.2 \\
07--0.275\dotfill &  0.0 &  0.0 &  0.0 &  0.0 &  0.1 &  0.7 & 18.2 &   0.0 &  0.0 &  0.0 &  0.0 &  0.0 &  0.0 &  0.0 &  0.0 &  0.0 &  0.0 &  0.0 &  0.0 &  0.0 &  0.0 \\
8--0.1\dotfill &  0.0 &  0.0 &  0.1 &  0.3 &  2.0 &  6.5 &  0.0 &  70.2 & 10.1 &  0.4 &  0.0 &  0.0 &  0.0 &  0.0 &  0.0 &  1.0 &  0.0 &  0.0 &  0.0 &  0.0 &  1.0 \\
9--0.3\dotfill &  0.1 &  0.0 &  0.1 &  0.0 &  0.3 &  2.0 & 36.4 &  10.0 & 81.4 &  5.5 &  0.2 &  1.0 &  0.6 &  0.4 &  0.0 &  0.6 &  0.5 &  2.2 &  0.6 &  0.0 &  1.7 \\
10--0.5\dotfill &  0.0 &  0.0 &  0.0 &  0.0 &  0.0 &  0.7 &  0.0 &  0.4 &  0.2 & 58.8 &  9.4 &  0.0 &  0.0 &  0.0 &  2.9 &  4.4 &  1.9 &  4.4 &  2.9 &  0.0 &  0.0 \\
11--0.7\dotfill &  0.1 &  0.0 &  0.0 &  0.0 &  0.0 &  0.0 &  0.0 &    0.0 &  0.0 &  7.7 & 62.6 & 16.0 &  0.4 &  0.2 &  2.1 &  1.0 &  1.3 &  6.6 & 11.8 &  0.0 &  0.0 \\
12--0.9\dotfill &  0.0 &  0.0 &  0.0 &  0.0 &  0.0 &  0.0 &  0.0 &    0.2 &  0.4 &  0.0 &  4.1 & 44.4 &  4.8 &  0.7 &  0.0 &  0.0 &  0.0 &  1.9 &  1.2 &  0.0 &  0.0 \\
13--1.1\dotfill &  0.0 &  0.1 &  0.0 &  0.0 &  0.0 &  0.0 &  4.5 &    0.0 &  0.7 &  0.0 &  0.2 & 26.5 & 70.2 & 41.2 &  1.0 &  0.2 &  0.3 &  1.9 &  0.6 &  0.0 &  0.0 \\
14--1.3\dotfill &  0.0 &  0.0 &  0.0 &  0.0 &  0.0 &  0.0 &  0.0 &    0.0 &  0.1 &  0.0 &  0.2 &  0.3 & 20.2 & 53.2 &  3.5 &  0.0 &  0.0 &  0.0 &  0.0 &  0.0 &  0.0 \\
15--1.5\dotfill &  0.0 &  0.0 &  0.0 &  0.0 &  0.0 &  0.0 &  0.0 &    1.1 &  2.2 &  8.7 &  4.9 &  0.3 &  1.4 &  2.5 & 79.6 & 22.9 &  9.3 &  3.0 &  0.0 &  0.0 &  0.0 \\
16--1.7\dotfill &  0.0 &  0.1 &  0.0 &  0.0 &  0.0 &  0.0 &  0.0 &    1.4 &  1.2 &  4.9 &  1.0 &  0.0 &  0.0 &  0.0 &  8.7 & 61.3 & 37.5 &  1.1 &  0.6 &  0.0 &  0.0 \\
17--1.9\dotfill &  0.0 &  0.0 &  0.0 &  0.0 &  0.0 &  0.0 &  0.0 &    0.4 &  0.3 &  0.9 &  0.0 &  0.0 &  0.0 &  0.0 &  0.8 &  7.7 & 42.6 &  9.9 &  0.0 &  0.0 &  0.7 \\
18--2.1\dotfill &  0.1 &  0.0 &  0.0 &  0.0 &  0.0 &  0.0 &  4.5 &    0.0 &  2.1 & 11.1 &  3.7 &  7.8 &  0.4 &  0.0 &  0.2 &  0.6 &  6.4 & 65.2 &  7.1 &  0.0 &  0.0 \\
19--2.3\dotfill &  0.0 &  0.0 &  0.0 &  0.0 &  0.0 &  0.0 &  0.0 &    0.0 &  0.0 &  0.4 &  5.7 &  2.3 &  0.0 &  0.0 &  0.0 &  0.0 &  0.0 &  3.3 & 65.9 & 38.9 &  0.0 \\
20--hiz qso \dotfill &  0.1 &  0.0 &  0.0 &  0.0 &  0.0 &  0.0 &    0.0 &  0.0 &  0.0 &  0.0 &  0.2 &  0.0 &  0.0 &  0.0 &  0.0 &  0.0 &  0.0 &  0.3 &  4.7 & 13.9 &  0.0 \\
21--red star\dotfill &  0.1 &  0.0 &  0.0 &  0.0 &  0.2 &  0.0 &    0.0 &  0.0 &  0.0 &  0.0 &  0.0 &  0.0 &  0.0 &  0.0 &  0.0 &  0.2 &  0.0 &  0.0 &  0.0 &  0.0 & 61.7 \\
\enddata
\tablecomments{The values of matrix $\xi_{ij}$ are  the percentage
of true class $i$ objects classified into class $j$. 
The matrix is from the spectroscopic sample of 20,253 objects in the bright
magnitude range (see Table 1) that were not used in training of the
classifier and, therefore, it validates the  classifier. 
The table header gives the class ID number, 1 through 21.  
The table first column gives both the class ID number and the
class name.  Classes 2 through 7 are the galaxy redshift classes, the
class name indicating the redshift value corresponding to the middle
of the respective redshift bin. 
Classes 8 through 20 are the AGN redshift classes. Class  ``hiz
qso'' covers the high-redshift range of the redshift distribution of
unresolved AGN. Four of the classifier galaxy redshift classes 
at  $z > 0.275$ contain a very small number of objects and are omitted 
in the table, while objects misclassified from other classes into them
are responsible for a small seeming discrepancy  of the column normalization
to 100\%.  
}
\end{deluxetable}

\begin{deluxetable}{lcc}
\tabletypesize{\small}
\tablewidth{0pt}
\tablecaption{\sc Galaxy redshift estimation 
by ClassX and other photometric methods 
\label{tnew}
}
\tablehead{
\multicolumn{2}{c}{ } & \multicolumn{1}{c}{Correct Classification ($3\sigma$)} \\
\colhead{Estimation Method} & \colhead{$\sigma_{\Delta z}$} & \multicolumn{1}{c}{(\%)} \\
}
\startdata
ClassX\dotfill & 0.0340 &  99.1 \\  
Polynomial\dotfill & 0.0318 &  98.0 \\
Nearest neighbor\dotfill & 0.0365 &  98.5 \\
Kd-tree\dotfill & 0.0254 &  98.4 \\ 
\enddata
\tablecomments{ClassX galaxy redshift estimation shown in Figure~\ref{f2} 
is compared with various photometric redshift estimators for galaxies 
in Table~1 of Csabai et al.\ (2003). The second column is the 
standard deviation, $\sigma_{\Delta z}
 =  (\langle(z_{\rm sp}-z_{\rm ph})^2\rangle)^{1/2}$, where
$z_{\rm ph}=z_{\rm{clx}}$ in the case of the ClassX method.
The third column is the percentage of redshifts correctly determined
within $3\sigma$, which in the case of ClassX 
approximately corresponds to the percentage of redshifts  
within the diagonal plus two adjacent redshift bins. 
}
\end{deluxetable}

\begin{deluxetable}{lrrrrr}
\tablewidth{0pt}
\tablecaption{{\sc ClassX Derived Content of the SDSS DR2 Photometric Catalog at Different Magnitude Limits}\label{t_content} }
\tablehead{
{} &  \multicolumn{4}{c}{Fraction\tablenotemark{a} \ (\%)} & {} \\
Magnitude Limit\tablenotemark{b} &  \multicolumn{4}{r}{\hrulefill} & $N_{tot}$\tablenotemark{c} \\
{} & 
star &  galaxy &  AGN &  red star & 
 {} 
}
\startdata
bright\dotfill & 89.58 &  7.92 &  1.07 &  1.43 &    $3.8 \times 10^6$  \\
intermediate\dotfill & 82.65 & 11.13 &  3.18 &  3.04 &    $6.4 \times 10^6$  \\
faint\dotfill &78.58 & 11.26 &  7.14 &  3.03 &    $7.0 \times 10^6$  \\
\enddata
\tablenotetext{a}{Percentage of class objects as derived  by ClassX from the bright, intermediate, and faint photometric samples.} 
\tablenotetext{b}{The magnitude constraints for each sample are given
in Table~1.}
\tablenotetext{c}{Number of the DR2 photometric  catalog objects satisfying the magnitude and photometric quality  constraints for the respective samples.}
\end{deluxetable}

\begin{deluxetable}{lccccccccr}
\tablewidth{0pt}
\tablecaption{{\sc Catalog of 91,847 Galaxies and Resolved AGN from SDSS DR2 Photometric Catalog classified with ClassX}\label{t-catalog} }
\tablehead{ nn\mbox{\hspace{1cm}} & RA & Dec & $u$ & $g$ & $r$ & $i$ & $z$ & redshift & object 
}
\startdata
     1\dotfill & 238.64369 &  $-1.01144$ & 20.59 & 19.29 & 18.14 & 17.64 & 17.35 & $0.15-0.20$ & galaxy \\
     2\dotfill & 238.80254 &  $-0.98485$ & 20.80 & 19.41 & 18.53 & 18.08 & 17.87 & $0.15-0.20$ & galaxy \\
     3\dotfill & 238.70859 &  $-0.93962$ & 19.30 & 18.46 & 17.81 & 17.44 & 17.19 & $0.00-0.20$ &    AGN \\
     4\dotfill & 238.82249 &  $-1.05858$ & 20.72 & 20.09 & 19.49 & 19.02 & 19.08 & $0.00-0.20$ &    AGN \\
     5\dotfill & 238.74800 &  $-0.88861$ & 20.05 & 18.81 & 18.47 & 18.20 & 18.08 & $0.00-0.05$ & galaxy \\
     6\dotfill & 238.94415 &  $-0.90889$ & 20.76 & 20.33 & 20.10 & 19.93 & 19.71 & $0.40-0.60$ &    AGN \\
     7\dotfill & 239.14249 &  $-0.88114$ & 20.43 & 19.31 & 18.65 & 18.30 & 18.08 & $0.10-0.15$ & galaxy \\
     8\dotfill & 239.33315 &  $-0.97546$ & 20.30 & 19.81 & 18.87 & 18.51 & 18.01 & $0.20-0.40$ &    AGN \\
     9\dotfill & 239.27873 &  $-1.02799$ & 20.36 & 20.02 & 19.95 & 19.63 & 19.41 & $0.40-0.60$ &    AGN \\
\enddata
\tablecomments{ The catalog in its entirety is available in 
the electronic version of the paper; the table displays only the first 
nine entries. 
The catalog is the result of classification of a sample of  100,000 SDSS DR2 resolved photometric objects, of which 91,847 objects were classified as  galaxies and AGN. The sample is constrained to the ``faint'' magnitude range (see text for details). Columns {\it u, g, r, i, z} are dereddened magnitudes. Column ``redshift'' gives the redshift range as determined by the classifier. Column ``object'' is the object type assigned by the classifier.}
\end{deluxetable}

\end{document}